\documentclass[12pt,a4paper,dvips]{article}
\usepackage{a4p}
\usepackage{cite,mcite}
\usepackage{graphicx}
\usepackage{physics }
\usepackage{l3_title,ifthen}
\journalname{Phys. Lett. B}
\date{November 6, 2000}
\preprint{2000-135}
\def\ra{\rightarrow }

\def\epem{\mbox{e}^+\mbox{e}^- }
\def\pip{\pi^+ }
\def\pim{\pi^- }

\def\kos{\mbox{K}^0_{\rm S}}
\def\k{\mbox{K}}      

\def\q2{\mbox{Q}^2 }
\def\NP{{ Nucl. Phys. }}
\def\PL{{ Phys. Lett. }}
\def\ZfP{{ Z. Phys. }}
\def\NIM{{ Nucl. Inst. Meth. }}
\def\PRep{{ Phys. Rep. }}
\def\PR{{ Phys. Rev. }}
\def\PRL{{ Phys. Rev. Lett. }}
\def\EURO{{ Eur. Phys. J. }}
\newlength{\capindent}
\newlength{\capwidth}
\newlength{\figwidth}
\newcommand{\icaption}[2][!*!,!]{\hspace*{\capindent}%
  \begin{minipage}{\capwidth}
    \ifthenelse{\equal{#1}{!*!,!}}%
      {\caption{#2}}%
      {\caption[#1]{#2}}
  \end{minipage}}
\begin{document}
\begin{titlepage}
\title{
K$^{\rm 0}_{\rm S}$K$^{\rm 0}_{\rm S}$ Final State in Two-Photon Collisions \\
and Implications for Glueballs  
}
\author{The L3 Collaboration}
%
%
\begin{abstract}
The $\kos\kos$ final state in two-photon collisions is studied with
the L3 detector at LEP.
The mass spectrum is dominated
by the formation of the f$_2\,\!\!\!'$(1525) tensor meson 
in the helicity-two state with a two-photon width times the branching
ratio into $\rm K\bar{K}$ of $76 \pm\, 6 \pm\, 11\eV$.
A clear signal for the formation of the f$_{\rm J}$(1710) 
is observed and it is found to be dominated by the 
spin-two helicity-two state.
No resonance is observed in the mass region around $2.2\GeV$ and an upper limit
of $1.4\eV$  at 95\% C.L. is derived for the two-photon width 
times the branching ratio into $\kos\kos$ 
for the glueball candidate $\xi(2230)$.

\vspace*{.9cm}
\centerline{\it Dedicated to the memory of Prof. Bianca Monteleoni}
\vspace*{-.9cm}
\end{abstract}
%
%
\submitted
\end{titlepage}

\section{Introduction}

 The formation of resonances in two-photon collisions is studied
via the process
$\epem\ra\epem\gamma^*\gamma^*$ $\ra\epem{\rm R}\ra\epem\kos\kos$,
where $\gamma^*$ is a virtual photon.
The outgoing electron and  positron carry nearly  the full 
beam energy and their transverse momenta are usually so small that
they are not detected. In this case the two photons are quasi-real.
 The cross section for this process
is given by the convolution of the QED calculable 
luminosity function $\cal{L}_{\gamma\gamma}$, giving the flux of
the photons, with a Breit-Wigner function. This leads to the
proportionality relation between the measured cross section and the
two-photon width $\Gamma_{\gamma \gamma}(\mbox{R})$ of the resonance R
\begin{equation}
\sigma(\epem\ra\epem\mbox{R})={\cal{K}}\cdot\Gamma_{\gamma \gamma}(\mbox{R}),
\label{eq:sig_k_Ggg}
\end{equation}
where the
proportionality factor ${\cal{K}}$ is evaluated by a
Monte Carlo integration. 

 The quantum numbers of the 
resonance must be compatible with the initial state of the two quasi-real
photons. A neutral, unflavoured meson with even charge conjugation, J$\neq$1
and helicity-zero ($\lambda$=0) or two ($\lambda$=2) can be formed.
In order to decay into $\kos\kos$, a resonance must have
J$^{PC}=(\mbox{even})^{++}$. 
For the $2^{++}$, 1$^3$P$_2$ tensor meson  nonet, the f$_2$(1270), the a$_2^0$(1320) 
and the  f$_2\,\!\!\!'$(1525) can be formed.
However, since these three states are close in mass,
interferences must be taken into account.
The f$_2$(1270) interferes constructively with the a$_2^0$(1320) in the
$\mbox{K}^+\mbox{K}^-$ final state but destructively 
in the $\k^0\bar{\k^0}$ final state~\cite{Lipkin}.

 Gluonium states cannot be formed directly by the collision
of two photons and the two-photon width of a glueball is expected to be
very small. A state that is observed in a gluon rich environment but
not in two photon fusion has the typical signature of a glueball.

 The data used for this analysis correspond
to an integrated luminosity of $588\,\pb$ collected by the L3 detector~\cite{L3} at LEP 
around the Z pole ($143\,\pb$) and at high energies, $\sqrt s=183-202\GeV$ ($445\,\pb$).
The $\kos\kos$ final state in two-photon collisions was studied 
by L3~\cite{L3-K0K0} with a luminosity of $114\,\pb$
and by TASSO, PLUTO and CELLO at lower energies and luminosities
at PETRA~\cite{PETRA}.

 The EGPC~\cite{Linde} Monte Carlo generator  is used to describe two-photon
resonance  formation. The generator is based on the formalism of Reference~\citen{Budnev}. 
All the generated events are passed through the full detector simulation program 
based on GEANT~\cite{GEANT} and GHEISHA~\cite{GHEISHA}
and are reconstructed following the same procedure used for the data.
Time dependent detector inefficiencies, as monitored during the data taking period,
are also simulated.

\section{Event selection}

 The selection of exclusive $\kos\kos$ events is based on the decay 
$\kos\ra\pip\pim$, exploiting the central tracking system and 
the electromagnetic calorimeter. 
The events  are collected 
predominantly by the charged particle track triggers~\cite{triggers}. 
In order to select $\epem\ra \epem\pip\pim\pip\pim$ 
events, we require:
\begin{itemize}
\item The total energy seen in the calorimeters must be smaller than $30\GeV$
to exclude annihilation events. 
\item There must be exactly four charged tracks in the tracking chamber with a net
charge of zero, 
a polar angle $\theta$ in the range $24^\circ<\theta<156^\circ$
and a transverse momentum greater than $100\MeV$. 
\end{itemize}

\begin{figure}[t]
\begin{center}
\includegraphics[height=.50\textwidth]{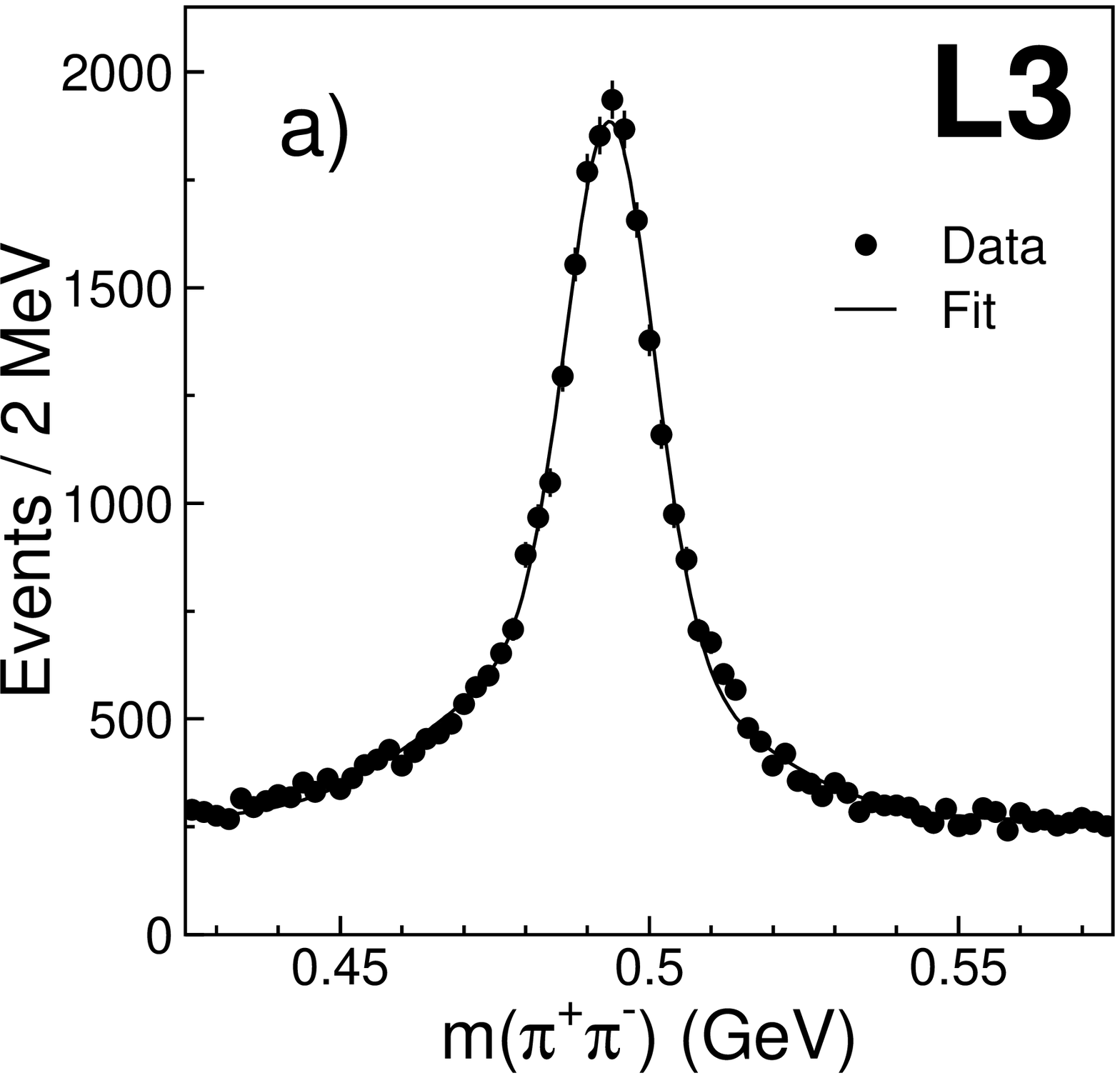}\includegraphics[height=.50\textwidth ]{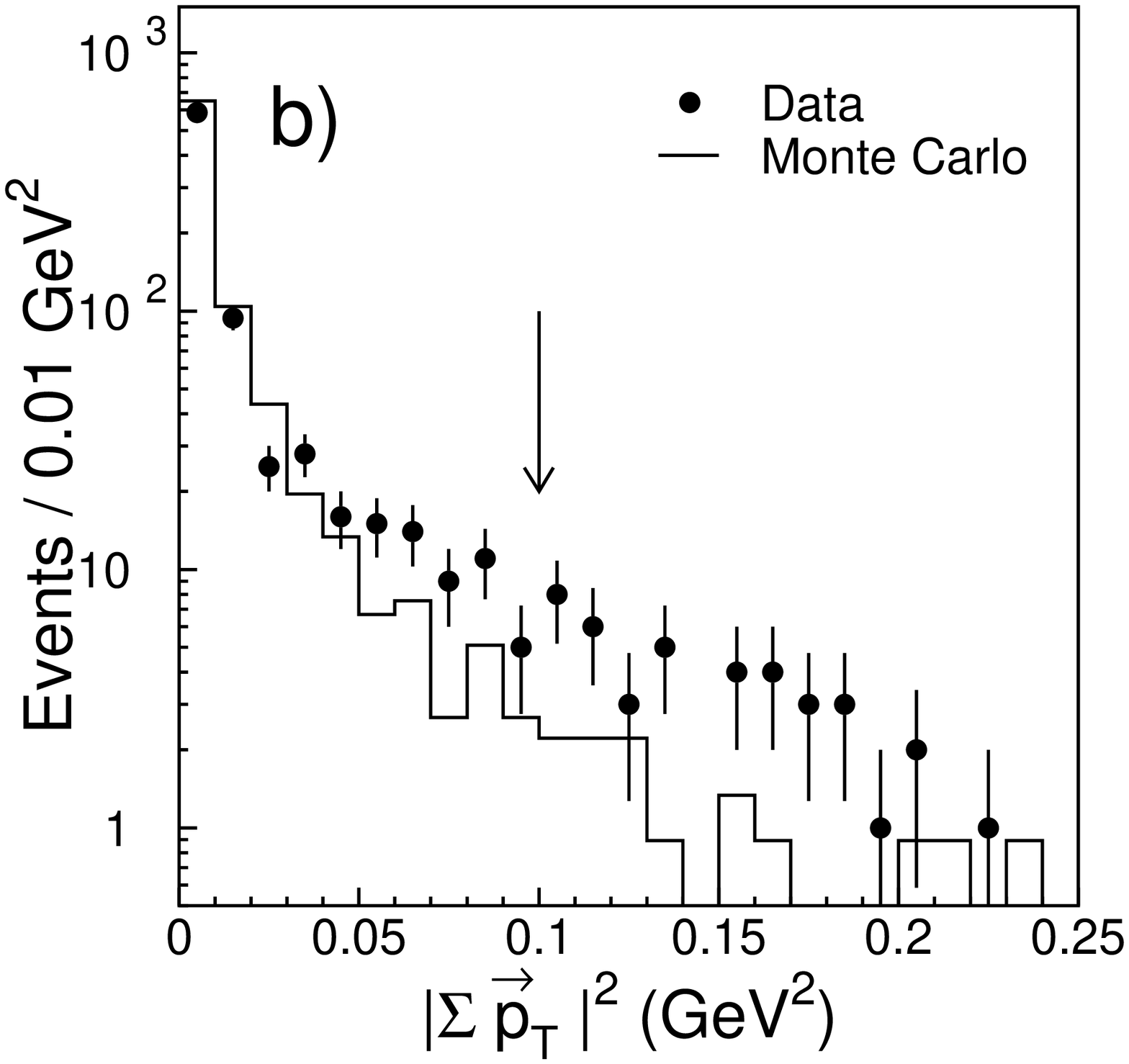}\\
\includegraphics[height=.50\textwidth]{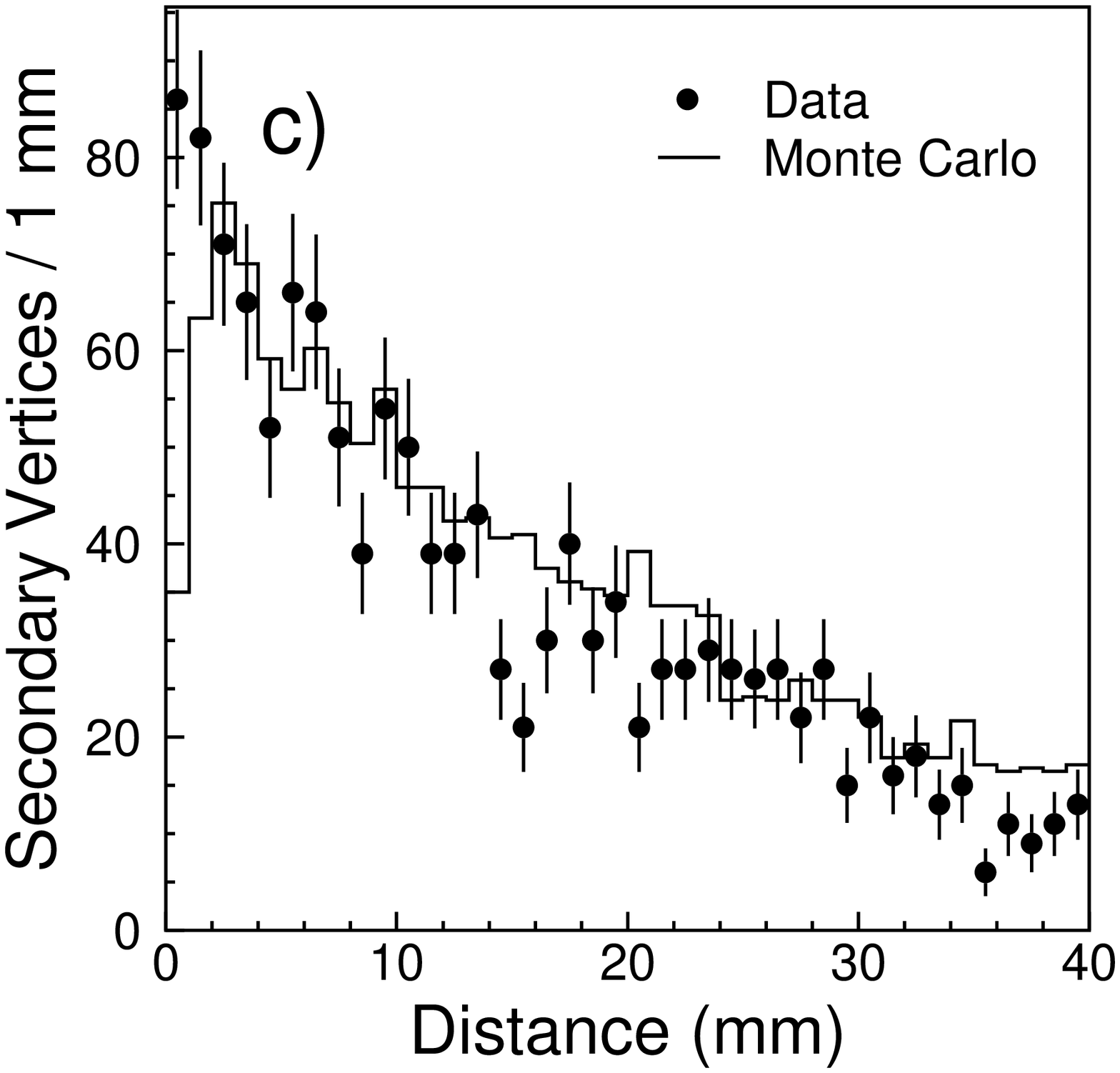}\includegraphics[height=.50\textwidth ]{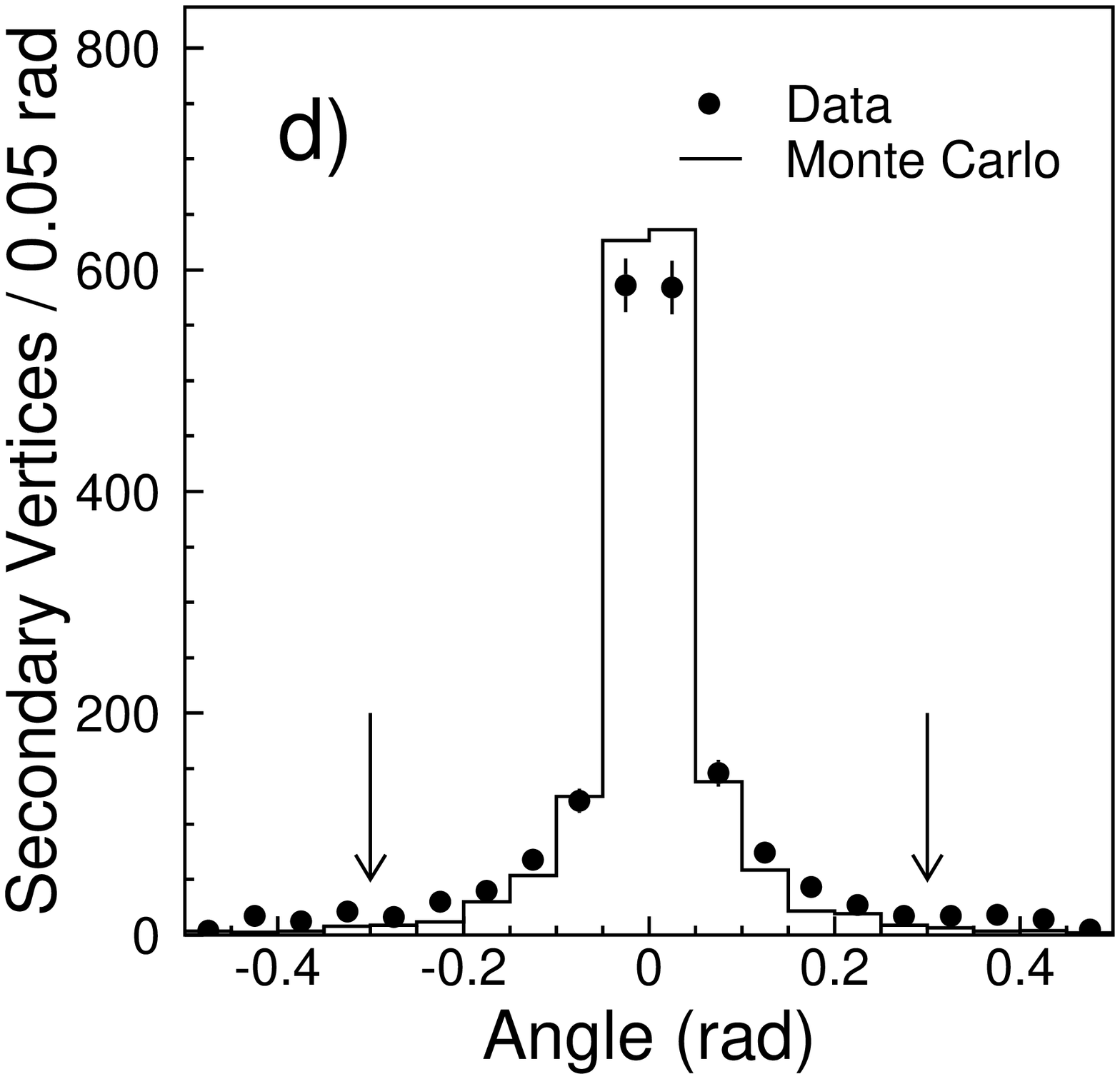}\\
\end{center}
\caption {
a)
The $\pip\pim$  mass spectrum for reconstructed secondary vertices 
with a transverse separation of
more than 3 mm from the interaction point.
b)
The total transverse momentum imbalance squared, 
c)
the distance between the
primary and 
the secondary vertex in the transverse plane  and 
d)
the angle between the flight direction and the total transverse momentum 
for the $\kos$ candidates. The Monte Carlo predictions correspond only to
the signal of $\kos\kos$ exclusive formation and 
are normalized to the same area as the data.
The arrows indicate the cuts applied.
}
\label{fig:k}
\end{figure}

The $\kos$'s are identified by secondary vertex reconstruction.
The $\pip\pim$ mass distribution is shown in
Figure~\ref{fig:k}a where a mass resolution
$\sigma=8.0\pm0.5\MeV$ is found, consistent with the Monte Carlo simulation.   

In order to select $\kos\kos$ exclusive events, we require:
\begin{itemize}
\item The total transverse momentum imbalance squared 
$|\sum \overrightarrow{p_T}|^2$
must be less than $0.1\GeV^2$.
In Figure~\ref{fig:k}b the $|\sum \overrightarrow{p_T}|^2$ distribution is
compared to the Monte Carlo prediction for exclusive $\kos\kos$ formation.
The excess in the data at high values is due to non-exclusive final states.
\item No photons.
A photon is defined as an isolated shower in the electromagnetic calorimeter
with a total energy larger than $100\MeV$ distributed in more than two crystals. 
The ratio between the energies deposited in the
hadronic and  electromagnetic  calorimeters  must be less than 0.2 and
there must be no charged track within
200 mrad from the shower direction. 
\item Two secondary vertices with transverse distances from the
interaction point greater than 1 mm and 3 mm.
In Figure~\ref{fig:k}c the data are
compared to the Monte Carlo prediction for exclusive $\kos\kos$ formation.
The excess in the data at low values is due to the dominant $\gamma\gamma\ra\rho^0\rho^0$
channel.
\item The angle between the flight direction of each $\kos$ candidate
and the total transverse momentum vector of the
two outgoing tracks in the transverse plane
must be less than 0.3 rad, as presented in Figure~\ref{fig:k}d.
\item Since the two $\kos$'s are produced back-to-back 
in the transverse plane, the angle between
the flight directions of the two $\kos$ candidates
in this plane is required to be $\pi\pm 0.3$ rad. 
\end{itemize}

\begin{figure}[t]
  \begin{center} 
   \includegraphics[height=1.0\textwidth ]{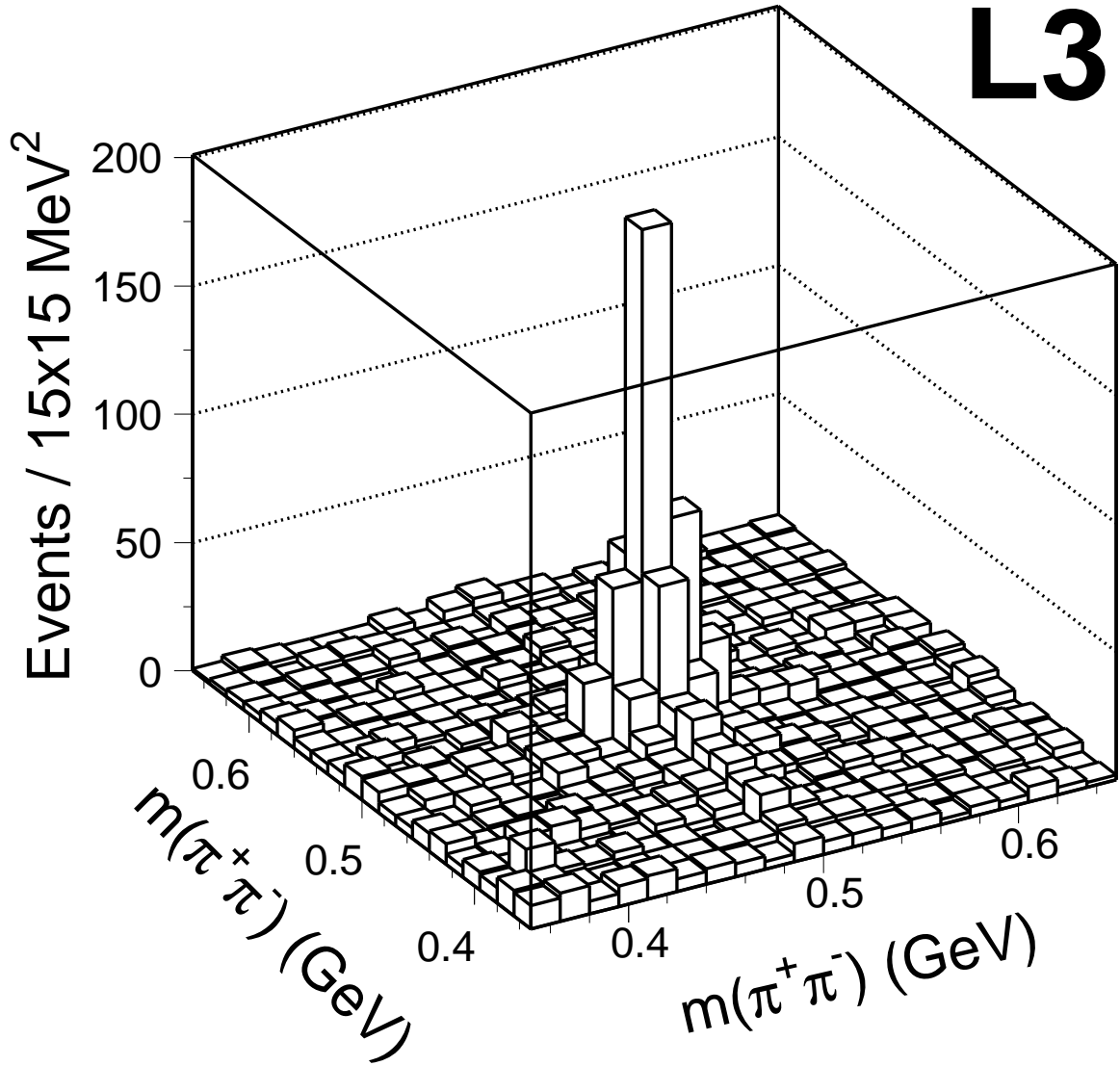}
\end{center}
\vspace*{-.7cm}
\caption{
The mass distribution of a $\kos$ candidate versus the mass of the other 
candidate for the full data sample.
A strong enhancement corresponding to the
$\kos\kos$ signal over a small background is observed. 
}
  \label{fig:kk}
\end{figure}

Figure~\ref{fig:kk} shows the distribution of the mass of
one $\kos$ candidate versus the mass of the other candidate.
There is a strong enhancement corresponding to the
$\kos\kos$ exclusive formation over a small background. We require
that the invariant masses of the two $\kos$ candidates must be inside
a circle of $40\MeV$ radius centred on the peak of the $\kos\kos$ signal.

With these selection criteria, 802 events are found in the data sample.
The background due to misidentified $\kos$ pairs and non-exclusive events
is estimated to be less
than 5\% by a study of the $\kos$ mass sidebands and of the 
$|\sum \overrightarrow{p_T}|^2$ distribution.  
The  backgrounds due to $\kos\k^\pm \pi^\mp$ and 
$\Lambda\bar{\Lambda}$ final states 
and to beam-gas and beam-wall interactions are found to be negligible.

\section{The $\kos\kos$ mass spectrum}

\begin{figure}[t]
  \begin{center}
   \includegraphics[height=1.0\textwidth ]{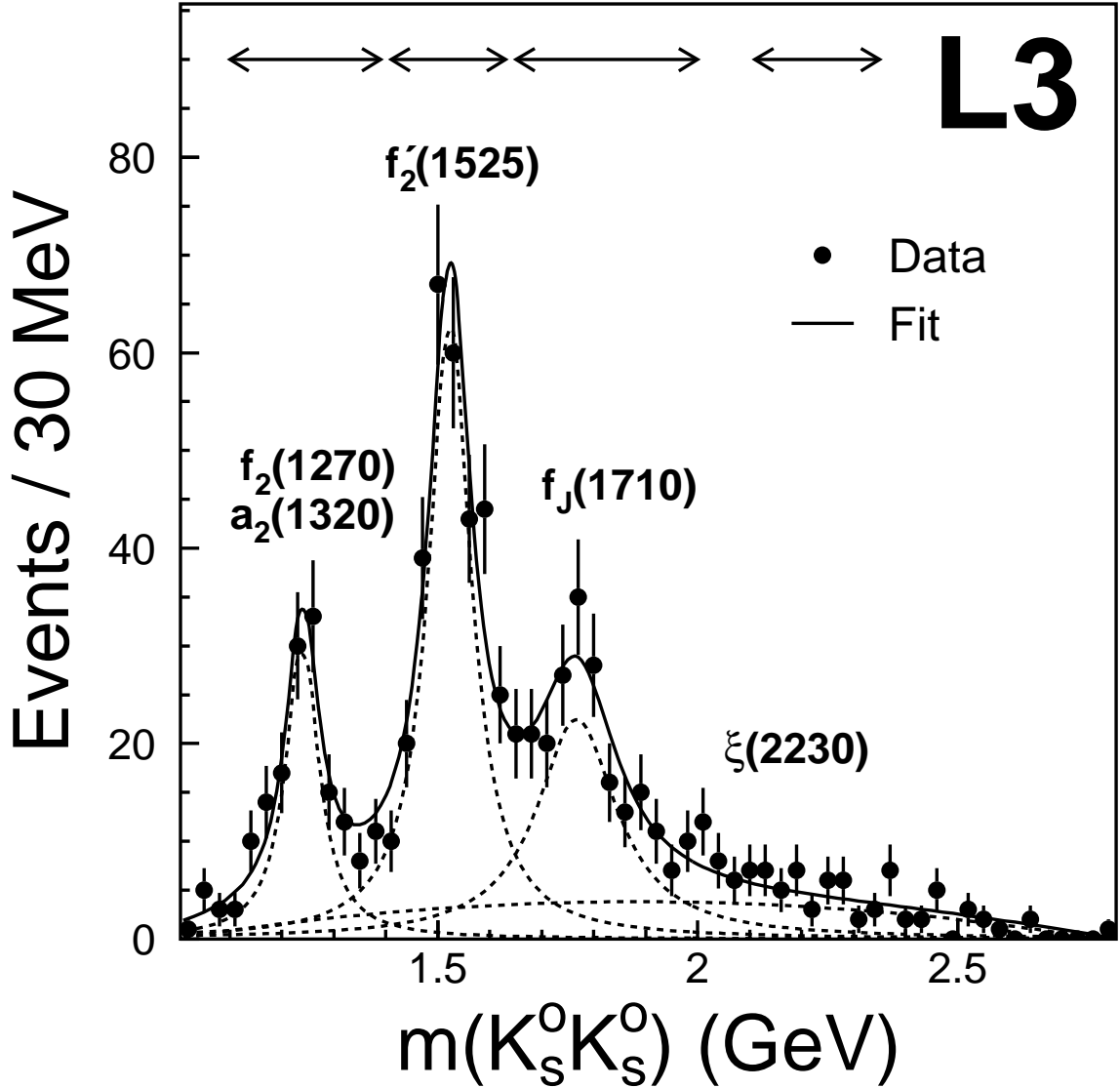}
  \end{center}
\vspace*{-.5cm}
\caption{
The $\kos\kos$ mass spectrum: the solid line
corresponds to the maximum likelihood fit. The background is
fitted by a second order polynomial and the three peaks by 
Breit-Wigner functions (dashed lines). The arrows correspond to the
f$_2$(1270)$-$a$_2^0$(1320), the f$_2\,\!\!\!'$(1525), the
f$_{\rm J}$(1710) and the $\xi$(2230) mass regions. 
}  
  \label{fig:kk_sp}
\end{figure}

 The $\kos\kos$ mass spectrum is presented in 
Figure~\ref{fig:kk_sp} showing three distinct peaks
over a low background.
Despite their large two-photon widths, the f$_2$(1270)
and the a$_2^0$(1320) tensor mesons produce a small signal
in the $\kos\kos$ final state due to their
destructive interference. 
The spectrum is dominated by the formation of the
f$_2\,\!\!\!'$(1525) tensor meson in agreement with previous
observations~\cite{L3-K0K0,PETRA}. 
A signal for the formation of the f$_{\rm J}$(1710) is 
present while no resonance is observed in the mass region of the
$\xi(2230)$.

 A maximum likelihood fit using three Breit-Wigner 
functions plus a second order polynomial for the background is 
performed on the full $\kos\kos$ mass spectrum. 
The results of the fit are shown in 
Figure~\ref{fig:kk_sp} and reported in 
Table~\ref{table:fit}. The confidence level is 31.7\%.
The parameters of the f$_2\,\!\!\!'$(1525) are in
good agreement with the PDG~\cite{PDG},
taking into account the experimental resolution $\sigma = 29 \pm 4\MeV$.

\section{The f$_2\,\!\!\!'$(1525) tensor meson}

 To study the f$_2\,\!\!\!'$(1525) tensor meson Monte Carlo events are
generated according to the mass, total width and two-photon width~\cite{PDG}
of this state.
The angular distribution of the two $\kos$'s in the two-photon centre of mass system
is generated uniformly in $\cos\theta^*$ and in $\phi^*$, 
the polar and azimuthal angles defined by the beam line.
In order to take into account the helicity of a spin-two resonance, 
a weight $w$ is assigned to each generated event:
$w=(\cos^2\theta^*-\frac{1}{3})^2$ for spin-two helicity-zero (J=2, $\lambda$=0) 
and
$w=\sin^4\theta^*$ for spin-two helicity-two (J=2, $\lambda$=2).

\begin{figure}[t]
  \begin{center}
   \includegraphics[height=1.0\textwidth ]{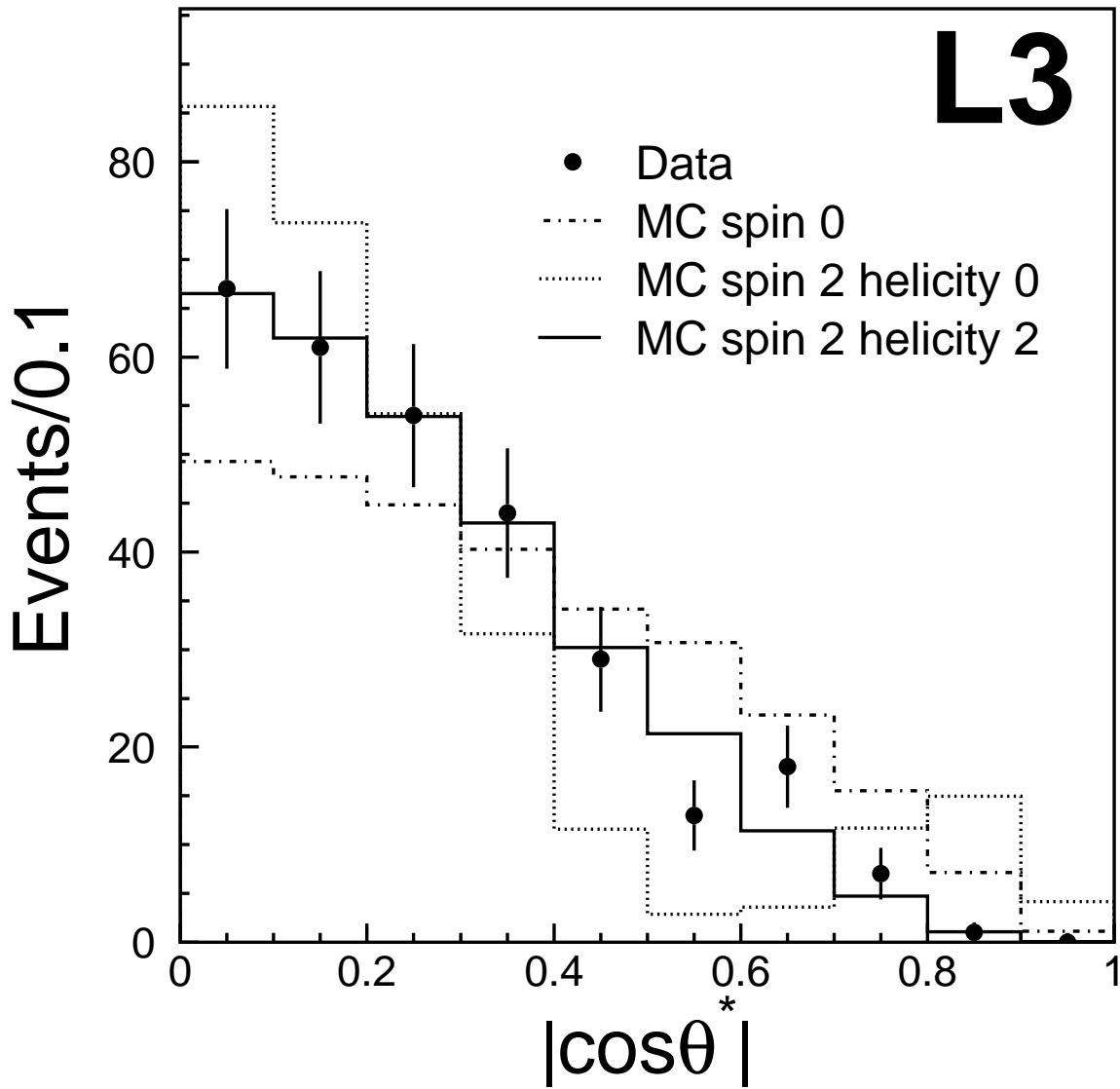}
  \end{center}
\vspace*{-.8cm}
\caption{
The $\kos\kos$ polar angle distribution 
compared with the Monte Carlo distributions for the hypothesis 
of a pure spin zero, spin-two helicity-zero and  spin-two helicity-two 
states for the f$_2\,\!\!\!'$(1525). 
The Monte Carlo expectations are normalized to the same number of events 
as the data. 
}
  \label{fig:ad_f2}
\end{figure}

 To determine the spin and the helicity composition in the f$_2\,\!\!\!'$(1525)
mass region between 1400 and $1640\MeV$,
the experimental polar angle distribution is compared with the
normalized Monte Carlo expectations for the (J=0),
(J=2, $\lambda$=0) and (J=2, $\lambda$=2) states, as presented in
Figure~\ref{fig:ad_f2}.
A $\chi^2$ is calculated
for each hypothesis, after grouping bins
in order to have at least 10 entries both in the data
and in the Monte Carlo.
The confidence levels for the (J=0) and (J=2, $\lambda$=0) 
hypotheses are less then $10^{-6}$. For the
(J=2, $\lambda$=2) hypothesis, a confidence level of 99.9\%
is obtained. The contributions of (J=0) and (J=2, $\lambda$=0) 
are found to be compatible with zero when fitting the three waves
simultaneously. The contribution of (J=2, $\lambda$=2) is found to be
compatible with unity within 7\%, in agreement with
theoretical predictions~\cite{Kopp}.

 The two-photon width times the branching ratio 
into $\k\bar{\k}$ 
is therefore determined from the cross section 
under the hypothesis of a pure (J=2, $\lambda$=2) state.
 Two separate measurements are performed for data collected 
at the Z pole and at high energies.
  The $\cal{K}$ factor,
the total detection efficiency and the measured quantity
$\Gamma_{\gamma\gamma}(\mbox{f}_2\,\!\!\!'(1525))\times
\mbox{Br(f}_2\,\!\!\!'(1525)\ra\k\bar{\k})$ are reported in
Table~\ref{table:f2}. 
The total detection efficiency, $\varepsilon$, is determined by Monte Carlo
and includes detector acceptance, trigger efficiency and selection criteria.
Combining the two measurements, the value
\begin{eqnarray*}
\Gamma_{\gamma\gamma}\Big(\mbox{f}_2\,\!\!\!'(1525)\Big)\times
\mbox{Br}\Big(\mbox{f}_2\,\!\!\!'(1525)\ra\k\bar{\k}\Big) = 76 \pm 6 \pm 11 \eV 
\end{eqnarray*}
is obtained, where the first uncertainty is statistical and the second is systematic.
The main source of systematic uncertainty
comes from the fitting procedure. A contribution of 10\% is evaluated by
varying the shape of the background and by allowing no background at all.
Other sources of systematic uncertainties are
trigger efficiency (5\%) and cut variations (7\%).
This result agrees with and supersedes the value
previously published by L3~\cite{L3-K0K0}.

\section{The \boldmath{1750\,\MeV} mass region}

 According to lattice QCD predictions~\cite{LatticeQCD}, the ground state glueball has
J$^{PC}$= 0$^{++}$ and a mass between 1400 and $1800\MeV$. 
Several 0$^{++}$ states are observed in this
mass region~\cite{PDG} and the scalar ground state glueball can  mix with nearby 
quarkonia~\cite{CMG}.

 To investigate the spin composition in the mass region of the
f$_{\rm J}$(1710), 
the angular distribution of the two $\kos$'s is studied in the
mass region between 1640 and $2000\MeV$. 
A resonance with a mass of $1750\MeV$ and
a total width of $200\MeV$ is generated as
for the f$_2\,\!\!\!'$(1525). The detection efficiencies
for the various spin and helicity hypotheses are reported in Table~\ref{table:1750}.

\begin{figure}[t]
  \begin{center}
    \includegraphics[height=1.0\textwidth ]{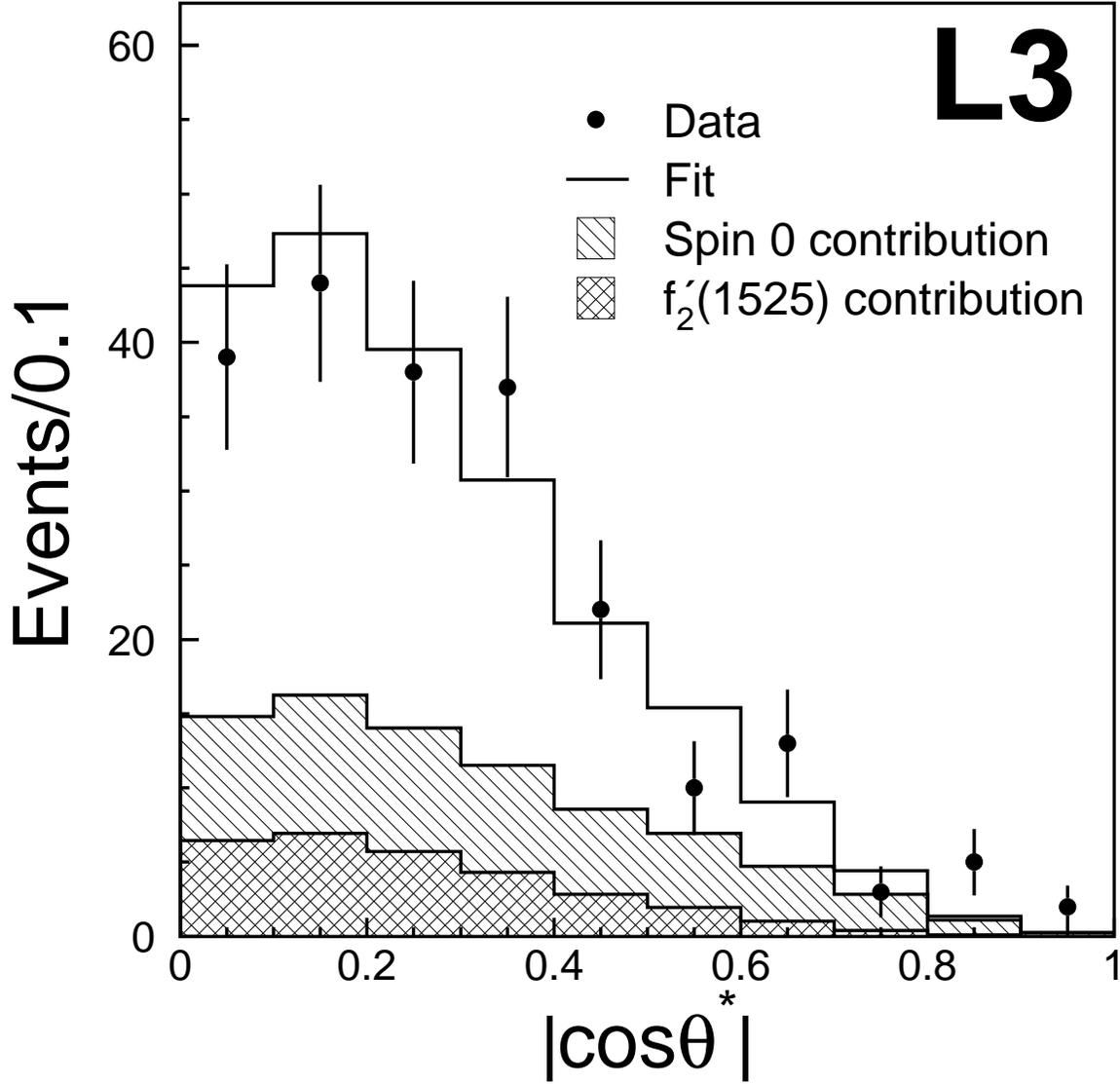}
  \end{center}
\vspace*{-.8cm}
\caption{
The fit of the $\kos\kos$ polar angle distribution
in the $1640-2000\MeV$ mass region. The contributions of spin-zero and
spin-two helicity-two waves are shown together with the 14\% contribution
of the tail of the f$_2\,\!\!\!'$(1525). 
}
  \label{fig:ad_18}
\end{figure}

 A fit of the angular distribution is performed using a combination of the two
waves (J=0) and (J=2, $\lambda$=2) for the signal plus the
distribution of the tail of the f$_2\,\!\!\!'$(1525).
Contributions from the (J=2, $\lambda$=0) wave are not considered,
based on the theoretical predictions~\cite{Kopp} and our experimental results for the 
f$_2\,\!\!\!'$(1525).
The tail of the f$_2\,\!\!\!'$(1525) is modeled by
assuming a pure (J=2, $\lambda$=2) state. 
The fraction of the events belonging to the f$_2\,\!\!\!'$(1525)
in the $1750\MeV$ mass region is found to be 14\%.
The fit results are shown in Figure~\ref{fig:ad_18}.
The confidence level for the fit is 68.0\%
whereas the (J=0) fraction is
24$\pm$16\%. Neglecting the (J=0) wave yields a confidence level of 24.0\%.
The possible (J=2, $\lambda$=0) contribution
is  found to be compatible with zero.
The (J=2, $\lambda$=2) wave is found to be dominant and
the two-photon width measured in the full data sample is
\begin{eqnarray*}
\Gamma_{\gamma\gamma}\Big(\mbox{f}_2(1750)\Big)\times
\mbox{Br}\Big(\mbox{f}_2(1750)\ra\k\bar{\k}\Big)= 49 \pm 11 \pm 13  \eV.
\end{eqnarray*}
The systematic uncertainty takes into account the selection criteria, the trigger efficiency,
the fitting procedure, the uncertainty on the total width and on the
(J=2, $\lambda$=2) fraction.

 The (J=2, $\lambda$=2) signal may be due to the formation of the first
radial excitation of a tensor meson state,
predicted at a mass of $1740\MeV$~\cite{Munz,Shakin}
with a two-photon width of $1.04\keV$~\cite{Munz}.
 The BES Collaboration reported the presence of both 2$^{++}$ and 0$^{++}$
waves in the $1750\MeV$ mass region in K$^+$K$^-$ in the
reaction $\epem\ra\mbox{J}\ra\mbox{K}^+\mbox{K}^-\gamma$~\cite{BESKPKM}. 
Their (J=0) fraction is
estimated to be 30$\pm$10\%, in good agreement with our measurement.

\section{The \boldmath{2230\,\MeV} mass region}

 The $\xi$(2230) is considered a good candidate for the ground state tensor
glueball because of its narrow width and its production in a gluon rich environment.
Its mass is in agreement with the lattice QCD predictions~\cite{LatticeQCD}.
It was first observed in the radiative decays of the J particle by
Mark III~\cite{MARKIII-xi} and confirmed by BES~\cite{BES-xi}.

 Since gluons do not couple directly  to photons, the two-photon width
is expected to be small for a glueball, as quantified by the 
stickiness~\cite{Chanowitz} defined as
\begin{equation}
\frac{|<{\rm R}|{\rm gg}>|^2}{|<{\rm R}|\gamma\gamma>|^2}\sim
S_{\rm R}=N_l \left(\frac{m_{\rm R}}{k_{J\rightarrow\gamma {\rm R}}}\right) ^{2l+1} 
\frac{\Gamma(\mbox{J}\rightarrow\gamma {\rm R})}{\Gamma({\rm R}\rightarrow\gamma\gamma)}
\label{eq:stickiness}
\end{equation}  
where $m_{\rm R}$ is the mass of the state R, $k_{J\rightarrow\gamma {\rm R}}$ is the energy of the photon from a
radiative J decay in the J rest frame and $l$ is the orbital angular momentum between the two gluons.
For spin-two states $l=0$. The normalization factor
$N_l$ is calculated assuming the stickiness of the f$_2$(1270) tensor meson to be 1. 

 A Monte Carlo simulation is used to
determine the detection efficiency for the $\xi$(2230) using
a mass of  $2230\MeV$ and a total width of $20\MeV$. A mass resolution
of $\sigma= 60\MeV$ is estimated. 
The total detection efficiency is reported in Table~\ref{table:csi} for the two data
samples,
under the hypothesis of a pure (J=2, $\lambda$=2) state.
The signal region is chosen to be $\pm2\sigma$ around the 
$\xi$(2230) mass. In order to evaluate the background two sidebands of $2\sigma$ 
each are considered.
The number of events in the signal region and the expected background 
evaluated with a  linear fit in the sideband region are reported in Table~\ref{table:csi}.
Using a Poisson distribution with background~\cite{Helene} and combining the
results for the two data samples,
we obtain the upper limit
\begin{eqnarray*}
\Gamma_{\gamma\gamma}\Big(\xi(2230)\Big)\times{\rm Br}\Big(\xi(2230)\ra\kos\kos\Big)<1.4 \eV
\rm{\,\,at \,\,95\%\,\, C.L.},
\end{eqnarray*}
under the hypothesis of a pure helicity-two state.

 This translates, following Equation~\ref{eq:stickiness} and using world average values~\cite{PDG},
into a lower limit on the stickiness $S_{\xi(2230)}$ of 74 at 95\% C.L.,
similar to the results obtained
by CLEO~\cite{CLEO}. This value of the stickiness is
much larger than the values measured for the well established $\rm q\bar{q}$ states
and supports the interpretation of the $\xi$(2230) as the tensor glueball. 

\section*{Acknowledgements}

We wish to 
express our gratitude to the CERN accelerator divisions for
the excellent performance of the LEP machine. 
We acknowledge the contributions of the engineers 
and technicians who have participated in the construction 
and maintenance of this experiment. 

%
\newpage
%


\newpage
\typeout{   }     
\typeout{Using author list for paper 223 -- ? }
\typeout{$Modified: Tue Sep  5 19:04:46 2000 by clare $}
\typeout{!!!!  This should only be used with document option a4p!!!!}
\typeout{   }
%
%
%
%
%
%

\newcount\tutecount  \tutecount=0
\def\tutenum#1{\global\advance\tutecount by 1 \xdef#1{\the\tutecount}}
\def\tute#1{$^{#1}$}
\tutenum\aachen            
\tutenum\nikhef            
\tutenum\mich              
\tutenum\lapp              
\tutenum\basel             
\tutenum\lsu               
\tutenum\beijing           
\tutenum\berlin            
\tutenum\bologna           
\tutenum\tata              
\tutenum\ne                
\tutenum\bucharest         
\tutenum\budapest          
\tutenum\mit               
\tutenum\debrecen          
\tutenum\florence          
\tutenum\cern              
\tutenum\wl                
\tutenum\geneva            
\tutenum\hefei             
\tutenum\seft              
\tutenum\lausanne          
\tutenum\lecce             
\tutenum\lyon              
\tutenum\madrid            
\tutenum\milan             
\tutenum\moscow            
\tutenum\naples            
\tutenum\cyprus            
\tutenum\nymegen           
\tutenum\caltech           
\tutenum\perugia           
\tutenum\cmu               
\tutenum\prince            
\tutenum\rome              
\tutenum\peters            
\tutenum\potenza           
\tutenum\salerno           
\tutenum\ucsd              
\tutenum\santiago          
\tutenum\sofia             
\tutenum\korea             
\tutenum\alabama           
\tutenum\utrecht           
\tutenum\purdue            
\tutenum\psinst            
\tutenum\zeuthen           
\tutenum\eth               
\tutenum\hamburg           
\tutenum\taiwan            
\tutenum\tsinghua          

{
\parskip=0pt
\noindent
{\bf The L3 Collaboration:}
\ifx\selectfont\undefined
 \baselineskip=10.8pt
 \baselineskip\baselinestretch\baselineskip
 \normalbaselineskip\baselineskip
 \ixpt
\else
 \fontsize{9}{10.8pt}\selectfont
\fi
\medskip
\tolerance=10000
\hbadness=5000
\raggedright
\hsize=162truemm\hoffset=0mm
\def\r{\rlap,}
\noindent

M.Acciarri\r\tute\milan\
P.Achard\r\tute\geneva\ 
O.Adriani\r\tute{\florence}\ 
M.Aguilar-Benitez\r\tute\madrid\ 
J.Alcaraz\r\tute\madrid\ 
G.Alemanni\r\tute\lausanne\
J.Allaby\r\tute\cern\
A.Aloisio\r\tute\naples\ 
M.G.Alviggi\r\tute\naples\
G.Ambrosi\r\tute\geneva\
% ...
\newpage
\begin{list}{A}{\itemsep=0pt plus 0pt minus 0pt\parsep=0pt plus 0pt minus 0pt
                \topsep=0pt plus 0pt minus 0pt}
\item[\aachen]
 I. Physikalisches Institut, RWTH, D-52056 Aachen, FRG$^{\S}$\\
 III. Physikalisches Institut, RWTH, D-52056 Aachen, FRG$^{\S}$
\item[\nikhef] National Institute for High Energy Physics, NIKHEF, 
     and University of Amsterdam, NL-1009 DB Amsterdam, The Netherlands
\item[\mich] University of Michigan, Ann Arbor, MI 48109, USA
\item[\lapp] Laboratoire d'Annecy-le-Vieux de Physique des Particules, 
     LAPP,IN2P3-CNRS, BP 110, F-74941 Annecy-le-Vieux CEDEX, France
\item[\basel] Institute of Physics, University of Basel, CH-4056 Basel,
     Switzerland
\item[\lsu] Louisiana State University, Baton Rouge, LA 70803, USA
\item[\beijing] Institute of High Energy Physics, IHEP, 
  100039 Beijing, China$^{\triangle}$ 
\item[\berlin] Humboldt University, D-10099 Berlin, FRG$^{\S}$
\item[\bologna] University of Bologna and INFN-Sezione di Bologna, 
     I-40126 Bologna, Italy
\item[\tata] Tata Institute of Fundamental Research, Bombay 400 005, India
\item[\ne] Northeastern University, Boston, MA 02115, USA
\item[\bucharest] Institute of Atomic Physics and University of Bucharest,
     R-76900 Bucharest, Romania
\item[\budapest] Central Research Institute for Physics of the 
     Hungarian Academy of Sciences, H-1525 Budapest 114, Hungary$^{\ddag}$
\item[\mit] Massachusetts Institute of Technology, Cambridge, MA 02139, USA
\item[\debrecen] KLTE-ATOMKI, H-4010 Debrecen, Hungary$^\P$
\item[\florence] INFN Sezione di Firenze and University of Florence, 
     I-50125 Florence, Italy
\item[\cern] European Laboratory for Particle Physics, CERN, 
     CH-1211 Geneva 23, Switzerland
\item[\wl] World Laboratory, FBLJA  Project, CH-1211 Geneva 23, Switzerland
\item[\geneva] University of Geneva, CH-1211 Geneva 4, Switzerland
\item[\hefei] Chinese University of Science and Technology, USTC,
      Hefei, Anhui 230 029, China$^{\triangle}$
\item[\lausanne] University of Lausanne, CH-1015 Lausanne, Switzerland
\item[\lecce] INFN-Sezione di Lecce and Universit\`a Degli Studi di Lecce,
     I-73100 Lecce, Italy
\item[\lyon] Institut de Physique Nucl\'eaire de Lyon, 
     IN2P3-CNRS,Universit\'e Claude Bernard, 
     F-69622 Villeurbanne, France
\item[\madrid] Centro de Investigaciones Energ{\'e}ticas, 
     Medioambientales y Tecnolog{\'\i}cas, CIEMAT, E-28040 Madrid,
     Spain${\flat}$ 
\item[\milan] INFN-Sezione di Milano, I-20133 Milan, Italy
\item[\moscow] Institute of Theoretical and Experimental Physics, ITEP, 
     Moscow, Russia
\item[\naples] INFN-Sezione di Napoli and University of Naples, 
     I-80125 Naples, Italy
\item[\cyprus] Department of Natural Sciences, University of Cyprus,
     Nicosia, Cyprus
\item[\nymegen] University of Nijmegen and NIKHEF, 
     NL-6525 ED Nijmegen, The Netherlands
\item[\caltech] California Institute of Technology, Pasadena, CA 91125, USA
\item[\perugia] INFN-Sezione di Perugia and Universit\`a Degli 
     Studi di Perugia, I-06100 Perugia, Italy   
\item[\cmu] Carnegie Mellon University, Pittsburgh, PA 15213, USA
\item[\prince] Princeton University, Princeton, NJ 08544, USA
\item[\rome] INFN-Sezione di Roma and University of Rome, ``La Sapienza",
     I-00185 Rome, Italy
\item[\peters] Nuclear Physics Institute, St. Petersburg, Russia
\item[\potenza] INFN-Sezione di Napoli and University of Potenza, 
     I-85100 Potenza, Italy
\item[\salerno] University and INFN, Salerno, I-84100 Salerno, Italy
\item[\ucsd] University of California, San Diego, CA 92093, USA
\item[\santiago] Dept. de Fisica de Particulas Elementales, Univ. de Santiago,
     E-15706 Santiago de Compostela, Spain
\item[\sofia] Bulgarian Academy of Sciences, Central Lab.~of 
     Mechatronics and Instrumentation, BU-1113 Sofia, Bulgaria
\item[\korea]  Laboratory of High Energy Physics, 
     Kyungpook National University, 702-701 Taegu, Republic of Korea
\item[\alabama] University of Alabama, Tuscaloosa, AL 35486, USA
\item[\utrecht] Utrecht University and NIKHEF, NL-3584 CB Utrecht, 
     The Netherlands
\item[\purdue] Purdue University, West Lafayette, IN 47907, USA
\item[\psinst] Paul Scherrer Institut, PSI, CH-5232 Villigen, Switzerland
\item[\zeuthen] DESY, D-15738 Zeuthen, 
     FRG
\item[\eth] Eidgen\"ossische Technische Hochschule, ETH Z\"urich,
     CH-8093 Z\"urich, Switzerland
\item[\hamburg] University of Hamburg, D-22761 Hamburg, FRG
\item[\taiwan] National Central University, Chung-Li, Taiwan, China
\item[\tsinghua] Department of Physics, National Tsing Hua University,
      Taiwan, China
\item[\S]  Supported by the German Bundesministerium 
        f\"ur Bildung, Wissenschaft, Forschung und Technologie
\item[\ddag] Supported by the Hungarian OTKA fund under contract
numbers T019181, F023259 and T024011.
\item[\P] Also supported by the Hungarian OTKA fund under contract
  numbers T22238 and T026178.
\item[$\flat$] Supported also by the Comisi\'on Interministerial de Ciencia y 
        Tecnolog{\'\i}a.
\item[$\sharp$] Also supported by CONICET and Universidad Nacional de La Plata,
        CC 67, 1900 La Plata, Argentina.
\item[$\diamondsuit$] Also supported by Panjab University, Chandigarh-160014, 
        India.
\item[$\triangle$] Supported by the National Natural Science
  Foundation of China.
\end{list}
}
\vfill


\newpage

\newpage

\begin{table}[t]
\begin{center}
\begin{tabular}{|lcccc|}\hline
\rule[.2in]{0.0in}{0.0in}Mass Region & f$_2$(1270)$-$a$_2$(1320) & f$_2\,\!\!\!'$(1525)  & f$_{\rm J}$(1710)  & Background \\ \hline\hline
Mass  ($MeV$)         & 1239  $\pm$   6  & 1523 $\pm$  6    & 1767 $\pm$ 14  & $-$ \\ 
Width ($MeV$)         &  $\;\;\;\:$$\;\,$78  $\pm$  19  &   $\;\;\;\,$100 $\pm$ 15    &   $\;\,$187 $\pm$ 60  & $-$ \\ 
Integral (Events)      &  $\;\;\;\,$123  $\pm$  22  &   $\;\;\;\,$331 $\pm$ 37    &   $\;\,$221 $\pm$ 55 & 149 $\pm$ 21 \\ \hline 
\end{tabular}
\caption{The parameters of the three Breit-Wigner functions and the parabolic
background from the fit on the $\kos\kos$ mass spectrum.}
\label{table:fit}
\end{center}
\end{table}

\begin{table}[t]
\begin{center}
\begin{tabular}{|lccccc|}\hline
\rule[.2in]{0.0in}{0.0in}                 & $\cal{L}(\pb)$   & $\cal{K}$ ($\pb/\keV$) & $\varepsilon$ (\%)& N(f$_2\,\!\!\!'$) & $\Gamma_{\gamma\gamma}(\mbox{f}_2\,\!\!\!')\times\mbox{Br}(\mbox{f}_2\,\!\!\!'\ra
\k\bar{\k}) $ ($\eV$)   \\ \hline\hline
Z Pole             & 143 &    605   & 5.0 $\pm$ 0.4 &  42.0 $\pm$   7.8  & $83\,\pm\,15\,\pm\,12$ \\
High Energies      & 445 &    845   & 6.4 $\pm$ 0.5 &   220 $\pm$  13    & $75\,\pm\,\,\,\,7\,\pm\,11$ \\ \hline
\end{tabular}
\caption{The measurement of the two-photon width of the f$_2\,\!\!\!'$(1525) for the two data samples.
N(f$_2\,\!\!\!'$) is the integral of the Breit-Wigner function 
in the $1400-1640\MeV$ mass region.  }
\label{table:f2}
\end{center}
\end{table}

\begin{table}[t]
\begin{center}
\begin{tabular}{|cccc|}\hline
\rule[.2in]{0.0in}{0.0in}                 & J=0  & J=2, $\lambda=2$ &  J=2, $\lambda=0$  \\ \hline\hline
Z Pole           &  6.0 $\pm$ 0.5\%   &  8.2 $\pm$ 0.7\%  &  4.1 $\pm$ 0.3\% \\ 
High Energies    &  6.3 $\pm$ 0.5\%   &  8.7 $\pm$ 0.7\%  &  4.5 $\pm$ 0.3\% \\ \hline
\end{tabular}
\caption{The total detection efficiency for the f$_{\rm J}$(1710)
for the various spin and helicity hypotheses. }
\label{table:1750}
\end{center}
\end{table}

\begin{table}[t]
\begin{center}
\begin{tabular}{|lcccc|}\hline
\rule[.2in]{0.0in}{0.0in}                 &  $\cal{K}$($\pb/\keV$)  & $\varepsilon$(\%) & $N_{ev}$ & $N_{bkg}$    \\ \hline\hline
Z Pole           &    161   & 16.6 $\pm$ 1.4   & $\;\,$6 &  $\;\,$4.9  \\ 
High Energies    &    230   & 14.2 $\pm$ 1.2   & 36 & 45.4  \\ \hline
\end{tabular}
\caption{The $\cal{K}$ factor, the detection efficiency $\varepsilon$, the number $N_{ev}$
of observed events
and the expected background $N_{bkg}$ for the $\xi$(2230).}
\label{table:csi}
\end{center}
\end{table}

\newpage

\newpage

\newpage

\newpage

\newpage

\end{document}